\documentclass[runningheads]{llncs}

\usepackage[T1]{fontenc}
\usepackage{graphicx}
\usepackage{hyperref}
\usepackage{color}

\newcommand{\MuDoC}[0]{\textsc{MuDoC}}

\newcommand{\TexDoC}[0]{\textsc{TexDoC}}
\newcommand{\Tex}[0]{\textsc{Tex}}
\newcommand{\DoC}[0]{\textsc{DoC}}

\newcommand{\DocSearch}[0]{\textsc{DocSearch}}
\newcommand{\Doc}[0]{\textsc{Doc}}
\newcommand{\Search}[0]{\textsc{Search}}

\usepackage{fancyhdr}

\fancypagestyle{firstpage}{
    \fancyhf{} 
    \fancyfoot[C]{Accepted to International Conference on Artificial Intelligence in Education (AIED) 2026, Seoul, Korea (June 29 - July 3, 2026)} 
    
}

\begin{document}

\title{Impact of Multimodal and Conversational AI on Learning Outcomes and Experience}
\titlerunning{Impact of Multimodal and Conversational AI on Learning}

\author{Karan Taneja\inst{1}\and
Anjali Singh\inst{2}\and
Ashok K. Goel\inst{1}
}

\authorrunning{K. Taneja et al.}

\institute{Georgia Institute of Technology, Atlanta GA 30332, USA \and
University of Texas at Austin, Austin TX 78712, USA \\
\email{\{karan.taneja,ashok.goel\}@cc.gatech.edu, anjali.singh@ischool.utexas.edu}}

\maketitle
\thispagestyle{firstpage}

\begin{abstract}

Multimodal Large Language Models (MLLMs) offer an opportunity to support multimedia learning through conversational systems grounded in educational content. 
However, while conversational AI is known to boost engagement, its impact on learning in visually-rich STEM domains remains under-explored. 
Moreover, there is limited understanding of how multimodality and conversationality jointly influence learning in generative AI systems.
This work reports findings from a randomized controlled online study ($N=124$) comparing three approaches to learning biology from textbook content: (1) a document-grounded conversational AI with interleaved \textit{text-and-image} responses (\MuDoC), (2) a document-grounded conversational AI with \textit{text-only} responses (\TexDoC), and (3) a textbook interface with semantic search and highlighting (\DocSearch).
Learners using \MuDoC\ achieved the highest post-test scores and reported the most positive learning experience.
Notably, while \TexDoC~was rated as significantly more engaging and easier to use than \DocSearch, it led to the lowest post-test scores, revealing a disconnect between student perceptions and learning outcomes.
Interpreted through the lens of the Cognitive Load Theory, these findings suggest that conversationality reduces extraneous load, while visual-verbal integration induced by multimodality increases germane load, leading to better learning outcomes.
When conversationality is not complemented by multimodality, reduced cognitive effort may instead inflate perceived understanding without improving learning outcomes. 

\keywords{Multimodal AI  \and Conversational Multimedia Learning \and RCT}

\end{abstract}

\section{Introduction}
\label{sec:introduction}

Multimedia learning integrates verbal and visual representations and has been shown to improve knowledge retention and transfer compared to learning from text alone \cite{mayer_multimedia_2002}.  
This is particularly important in visually rich STEM disciplines, where visual representations such as diagrams and graphs are central to learning. 

Recent advances in Multimodal Large Language Models (MLLMs) create new opportunities to support learning by enabling conversational systems grounded in educational content that can generate personalized multimedia explanations in response to learner inquiries \cite{taneja_towards_2025}. 
This potential is especially salient given the rapid adoption of AI tools for learning and tutoring, including 
large-scale
deployments in higher education \cite{kakar_jill_2024}. 
However, empirical evidence regarding the educational impact of AI chatbots remains mixed. 
While some studies report improvements in learning outcomes and student engagement \cite{wu_ai_2024}, others raise concerns such as overreliance and reduced metacognitive effort \cite{zhai_effects_2024}.
Additionally, the impact of AI-generated multimedia content on learning outcomes is still not well understood. 
Finally, there is limited work examining how conversational interaction and multimodality jointly influence the learning process.

To investigate how multimodality and conversational AI---both independently and jointly---influence students' learning outcomes and experience, we conducted a randomized controlled trial (RCT) on Prolific with $N=124$ participants learning meiosis cell division, a fundamental topic in cell biology. 
We compared three systems: 
(i) \textbf{\MuDoC~2.0} (henceforth referred to as \MuDoC), an AI system for \textit{conversational multimedia learning} that uses text and visuals from learning material to construct interleaved text-and-image responses, extending our previous system \textbf{MuDoC 1.0} \cite{taneja_mudoc_2025},
(ii) \textbf{\TexDoC}, a system that is identical to \MuDoC, except that it generates text-only responses without images, and 
(iii) \textbf{\DocSearch} ({\Doc}ument \Search), a semantic search tool that supports multimedia learning by highlighting relevant content in the textbook in response to students' search queries.
Participants first completed a pre-test, then learned the assigned topic using one of the three systems, and finally completed a post-test and a survey probing their learning experience.

Results show that \MuDoC~led to significantly higher post-test scores compared to \TexDoC, whereas no significant differences were observed between the remaining pairs of systems. For learning experience, \MuDoC~and \TexDoC~were rated similarly, while \DocSearch~received significantly lower ratings.
Interpreting these findings from the lens of the Cognitive Load Theory (CLT) \cite{sweller_cognitive_2011} suggests that conversational multimedia learning supports learning by reducing extraneous load through conversationality while increasing peda-gogically-effective germane load through visual-verbal integration. 
On the other hand, conversational interaction without accompanying visuals enhances perceived learning experience despite resulting in poorer learning outcomes. 
These findings provide evidence in support of using MLLMs for conversational multimedia learning to enhance both students' learning outcomes and experience. Further, by examining learner behaviors and the disconnect between subjective satisfaction and objective performance of the \TexDoC~group, this work offers a new perspective on the effects of conversational AI on learning and over-reliance.

\section{Related Work}
\label{sec:related-work}

Cognitive Theory of Multimedia Learning (CTML) \cite{mayer_multimedia_2002} posits that learners process verbal and visual information through separate, capacity-limited channels.
Learning is enhanced when these representations are meaningfully integrated with one another and with prior knowledge in long-term memory \cite{mayer_multimedia_2002}. 
This integration supports deeper cognitive processing and leads to improved learning outcomes compared to verbal information alone.

Several text-based tutors have been used for learning and management in classrooms.
Jill Watson \cite{goel_jill_2018} was an early question-answering system for answering student queries on classroom forums based on structured information.
More recently, it utilizes LLM-based methods with retrieval-augmented generation \cite{taneja_jill_2024} and has been deployed at large scale at multiple institutions \cite{kakar_jill_2024}.   
Since the publication of this paper, MuDoC has been integrated into Jill Watson and has been deployed in classrooms at Georgia Institute of Technology.
\textit{JeepyTA} \cite{baker_step_2024} is another such virtual teaching assistant that uses LLM-generated responses and a list of question-answer pairs which are ranked based on similarity matching. 
\textit{Pedagogical Tutor} \cite{wolfel_knowledge-based_2024} similarly relies on template-based slides to answer questions with LLMs, and also allows users to search relevant slides.

Multimodal AI systems have also been explored for multimedia learning in several domains \cite{bland_enhancing_2025,chen_interactive_2025,chen_automatic_2024,taneja_towards_2025}, but the impact of AI-generated multimedia content on learning outcomes remains unclear.
Chen et al. \cite{chen_automatic_2024} generated visualizations as learning aids for poetry but did not compare against a non-visual baseline. 
Bland et al. \cite{bland_enhancing_2025} used AI to generate cinematic clinical narratives for pharmacology but focused only on student experience. 
In mathematics, the \textit{Interactive Sketchpad} \cite{chen_interactive_2025} tool uses an MLLM to generate and manipulate diagrams within a conversational tutoring setting. While learners reported a better learning experience and problem-solving efficacy using this tool compared to using ChatGPT, this study did not examine the impacts on their learning outcomes.
\textit{MuDoC 1.0} \cite{taneja_towards_2025}
enhanced learner engagement and trust but did not significantly improve learning outcomes compared to a baseline system with text-only responses.
To the best of our knowledge, prior work does provide evidence for the effectiveness of MLLMs in generating multimedia content for improving learning outcomes.

Regarding the effects of conversational AI, recent studies with educational chatbots, such as ChatGPT-style systems, report gains in academic performance, engagement, and perceived personalization, though effects vary across domains and contexts \cite{deng2025does,wu_ai_2024}.
At the same time, researchers have also raised concerns about overreliance and shallow processing due to undermined cognitive and metacognitive processes that are essential for durable learning \cite{fan_beware_2024}.
Further, it is unclear how conversational interfaces for interacting with textbooks compare to more classical document or semantic search baselines, which we address in this work.

\section{\MuDoC, \TexDoC, and \DocSearch}
\label{sec:MuDoC}

In this section, we first describe \MuDoC, an AI system that answers queries with interleaved text-and-image responses, extending MuDoC 1.0 \cite{taneja_mudoc_2025}, using the `Reason-and-Act' (ReAct) framework \cite{yao_react_2022} (see Figure \ref{fig:MuDoC}).
We then discuss \TexDoC~and \DocSearch~made with minimal modifications to the \MuDoC~pipe-line to isolate the effects of visuals and conversational AI on learning. 
Additional details about these AI systems are provided in the supplementary material\footnote{Supplementary material (including demo videos): \href{https://tinyurl.com/IMCAILOE}{https://tinyurl.com/IMCAILOE}\label{footnote:supplementary-material}}.

\subsection{Preprocessing and Response Generation in \MuDoC}
\paragraph{Document Preprocessing:} 
The document processing pipeline involves parsing the document layout to extract content from each page, which is subsequently processed and stored in a vector database to facilitate the retrieval process.
We utilize the HURIDOCS document layout analysis parser \cite{noauthor_huridocspdf-document-layout-analysis_2025} to extract text, figures, and other structural elements from the document. 
These \textit{blocks} extracted during the layout analysis are combined into partly overlapping \textit{chunks} of at least 8,000 characters.
For each text chunk, we generate and store a succinct summary using the OpenAI \texttt{gpt-4o-mini-2024-07-18} model\footref{footnote:supplementary-material}. 
Similarly, for every extracted image, we store a caption and a detailed image description generated using the same model.
We employ the OpenAI \texttt{text-embedding-3-large} model for text embeddings and 
the \texttt{google/siglip-so400m-patch14-384} model for image embeddings \cite{zhai_sigmoid_2023}. 
The resulting chunks and their embeddings are stored in a vector database for retrieval (described later).
Compared to MuDoC 1.0, among other differences, we now use a more recent and performant layout parser and text and image embedding models along with larger context windows.

\begin{figure}[t]
    \centering
    \includegraphics[width=\linewidth]{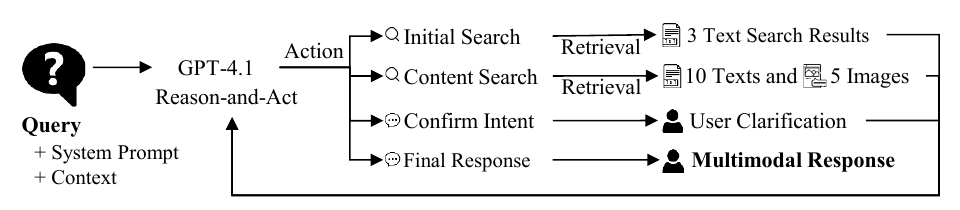}
    \caption{\MuDoC~uses GPT-4.1 for Reason-and-Act (ReAct) loop with 4 possible actions viz. Initial Search, Content Search, Confirm Intent, and Final Response. }
    \label{fig:MuDoC}
\end{figure}

\paragraph{Response Generation:}
\MuDoC~uses OpenAI \texttt{gpt-4.1-2025-04-14}, which supports multimodal inputs and function calling capabilities, for response generation. 
It~now uses reasoning, unlike MuDoC 1.0, by structuring response generation as a ReAct \cite{yao_react_2022} loop to improve search functions usage and generation quality. 
In each iteration of the ReAct loop, the model first \underline{Re}asons its current state within the conversation, 
and then decides to take an \underline{Act}ion to search for additional context or to generate a response (see Figure \ref{fig:MuDoC}).

\paragraph{Reasoning:} 
Prior to executing an action, the LLM is instructed to perform the following three reflection steps\footref{footnote:supplementary-material}.
    (i) \textbf{Query Reflection:} The model evaluates the user's intent, identifying any unknown terminology, acronyms, typos, or misconceptions within the query, and determines whether a search is necessitated by the request. 
    (ii) \textbf{Search Content Reflection:} If content has been retrieved during the current or past conversation turns, the model assesses content relevance and identifies any missing information that requires further retrieval. 
    The model also reasons about the most pertinent textual and visual content and determines how to structure a self-contained response, including definitions for terms that may be unfamiliar to the student.
    (iii) \textbf{Action Reasoning:} In this final step, the model synthesizes the previous reflections to select the most appropriate course of action. 
    The possible actions are detailed next.

\paragraph{Actions:}
The model can execute four distinct actions, categorized into search and generation functions. 
If the model requires foundational information to interpret the query, such as definitions for unfamiliar terms or acronyms, it utilizes the \textbf{Initial Search} action. 
This retrieves the top three text results, ensuring a low-overhead clarification of the problem space.
Conversely, if the model understands the query sufficiently but requires a search, it can trigger the \textbf{Content Search} action and
subsequently generate specific search queries for text and image retrieval function calls, which can return up to ten text chunks and five images respectively. 
The remaining two actions are generative.
If the model determines that the user's intent is ambiguous, it utilizes the \textbf{Confirm Intent} action to request a clarification, ensuring a precise final output. 
Finally, once the model possesses a clear understanding of the query and the necessary retrieved context, it executes the \textbf{Final Response} action to synthesize a comprehensive response for the user.
In the event of a search action, the retrieved results are appended to the conversation history, serving as updated context for the subsequent ReAct loop. 
Conversely, when a generation action is performed, the model delivers the response and awaits the user's input to proceed.

\paragraph{Text and Image Retrieval:}
As opposed to purely embedding-based retrieval in MuDoC 1.0, \MuDoC~now employs a hybrid search strategy \cite{karpukhin_dense_2020} combining dense vector embeddings with sparse keyword-based retrieval for both texts and images.
For text retrieval, we utilize both the original content and the generated summaries (from preprocessing) for embeddings and keyword-based scoring. 
We found that prioritizing vector embedding similarity (75\% weight) over keyword-based BM25 \cite{robertson_okapi_1994} score (25\% weight) led to the highest retrieval accuracy in hybrid ranking. 
After text retrieval, we sort retrieved results in the natural order of the textbook and remove redundancy caused by overlapping chunks, leading to a leaner and organized context for the LLM.
In image retrieval, the raw text index includes the image captions and descriptions while the image vector is the mean of the image and caption embedding. 
For ranking images, we found that prioritizing BM25 score (75\% weight) led to better retrieval, potentially because image descriptions focus on key terminology, where keyword matching works well, rather than semantically-rich explanations.
We plan to release a thorough evaluation of text and image retrieval methods used in \MuDoC~in future work.

\paragraph{Interleaved Grounded Generation:}
\MuDoC~synthesizes an interleaved, grounded response based on retrieved texts and images while providing explicit references to the source material. 
To incorporate visuals, the model is instructed to generate HTML \texttt{figure}, \texttt{figcaption} and \texttt{img} tags, based on source links and corresponding raw retrieved images.
Textual claims are supported by in-text citations, similar to WebGPT \cite{nakano_webgpt_2022}, referring to the original document name and block IDs to ensure that the generated response is grounded and verifiable.
We found that block-based in-text citations (using LLMs directly) had higher reliability than embedding-based source attribution used in MuDoC 1.0.

\paragraph{Pedagogically-aligned Prompt:} 
The model's generation is guided by the principles of CTML \cite{mayer_multimedia_2002}. 
Specifically, the model is prompted to prioritize functional visuals over decorative ones to minimize cognitive load.
Information provided via images should be complementary rather than redundant to the text, and visual elements should be spatially integrated near the relevant text.
The model is instructed to utilize concrete examples and analogies rather than abstract definitions in accordance with CTML's worked example effect.
Further, explanations should follow a `simple-to-formal' progression, grounding new concepts in the context of prior conversation turns.
At the conclusion of a response, \MuDoC~can pose reflective questions to encourage active learning, invite the user to explain concepts in their own words or make follow-up inquiries. 
The system prompt\footref{footnote:supplementary-material} also includes instructions to ensure that all responses remain safe, polite, and constructive.

\paragraph{Response Post-processing and Rendering:}
Following response generation, we employ regular expressions (regex) to identify in-text citations and image placeholders within the model's output. 
These identifiers are dynamically replaced during token streaming with interactive link icons that connect directly to the source document. 
When a user clicks a citation link, the system retrieves the precise coordinates of the target block from the backend, triggering the frontend to switch to the `Document' tab, navigate to the corresponding page, and highlight the relevant passage for several seconds to facilitate verification.
For HTML figure references, the specific image---extracted from the source document during document preprocessing---is embedded directly into the chat message.
Similar to text citations, each image is also accompanied by a link icon that allows the user to locate the figure within the original document. 
To ensure transparency, the model's reasoning process is accessible via a dedicated icon adjacent to the agent's avatar. 
Furthermore, the token stream is monitored in real-time to detect ReAct loops, allowing the interface to display agent status indicators when the system is performing reasoning and retrieval.

\subsection{\TexDoC~and \DocSearch}

\TexDoC, a {\Tex}t-only version of Mu\DoC, lacks the image retrieval pipeline of \MuDoC, 
image-related prompt instructions,
the image search function call,
and raw input images.
As a result, it provides text-only responses similar to Jill Watson \cite{taneja_jill_2024}, but with interactive citations, indirectly allowing users to find visuals in the textbook.
Aside from the omissions, \TexDoC~is identical to \MuDoC~and helps us isolate the impact of multimodality in our experiments.

\DocSearch~is a stateless semantic search tool integrated within the document viewer.
It employs the same retrieval pipeline as \TexDoC~and uses the same LLM to select the most relevant text and image blocks from the retrieved content to provide up to ten search results.
Users can navigate these results using arrow keys or by clicking corresponding buttons. 
As a user navigates to a search result, a block of text or image is highlighted for a few seconds, similar to \MuDoC~and \TexDoC~citations. 
Overall, \DocSearch~is a convenient way to navigate a document with very specific search queries, but it does not synthesize responses or use past context for follow-up questions, 
helping us study the impact of conversational AI compared to non-conversational multimedia.

\section{Experimental Methods} 
\label{sec:experiments-and-results}

\begin{figure}[t]
    \centering
    \includegraphics[width=\linewidth]{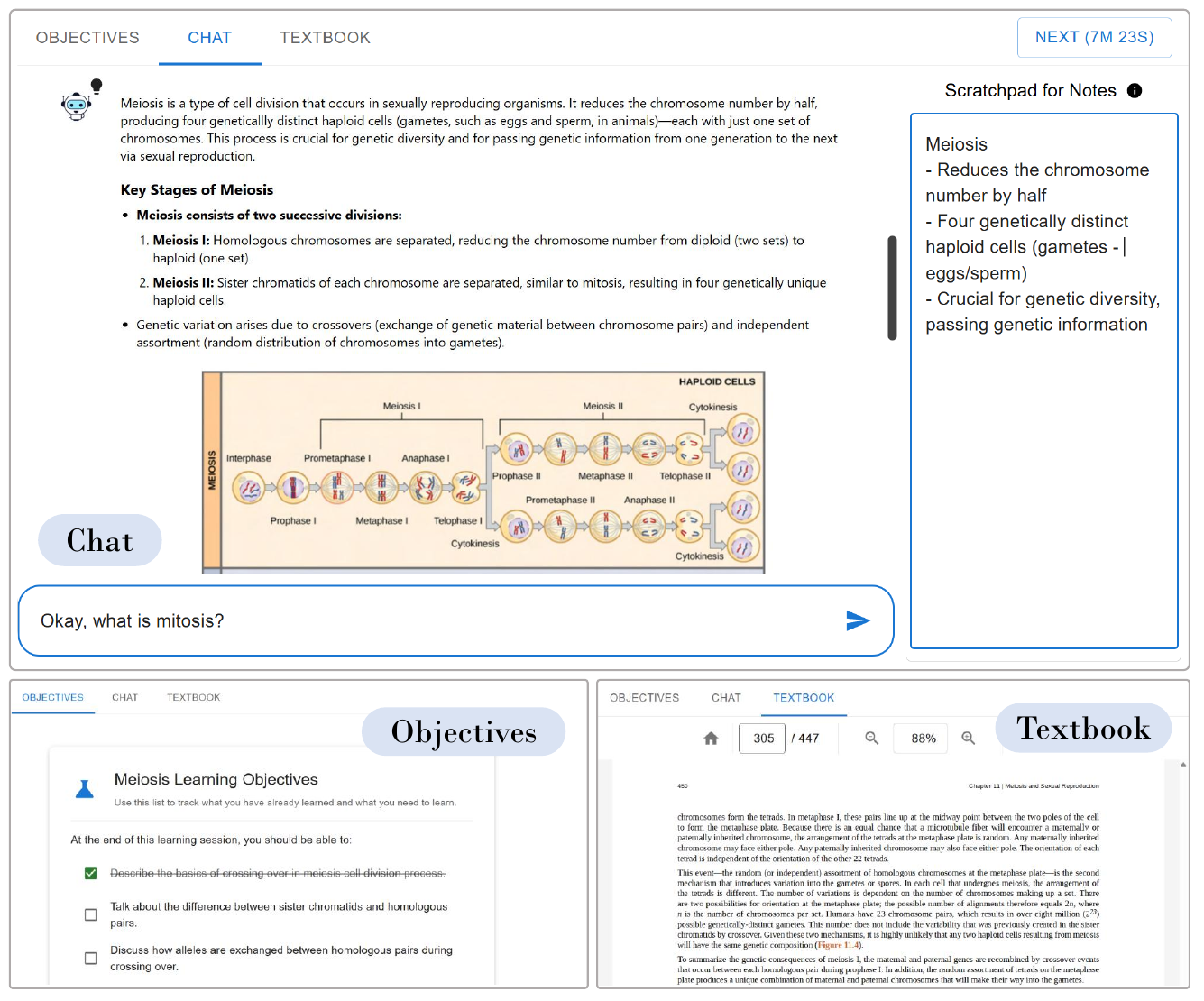}
    \caption{\textbf{Snapshots from self-learning session}: `Chat' tab shows the conversational AI interface (\MuDoC~and \TexDoC). `Objectives' tab contains the learning objectives with checkboxes to track progress. `Document' tab shows the source document for all AI systems. A `Notepad' box is present on the right for note-taking.}
    \label{fig:learning-session}
\end{figure}

\textbf{Participant Recruitment:}
We conducted the study using the online platform \href{https://www.prolific.com/}{Prolific}, recruiting participants aged 18+, residing in the US, who had a major in biology and had completed at least a high school diploma.
The compensation was US\$20 for completing the study with an estimated duration of 75 minutes.

A total of 162 participants completed the study, of which 17 were excluded for not passing attention checks, 18 were excluded for insufficient activity during the learning session, 
and 3 were excluded due to technical issues, resulting in a final sample size of $N=124$. Out of 124, 117 (94\%) of participants were 18-35 years old, and 112 (90\%) had basic familiarity with AI as end users.

\vspace{3pt}
\textbf{Study Protocol:}
The participants first signed the consent form for our study, approved by the Institutional Review Board (IRB) at Georgia Tech. 
Following consent, they were randomly assigned to one of the three experimental conditions, viz. \MuDoC, \TexDoC~and \DocSearch.
In each condition, participants first completed a pre-test with 10 multiple-choice questions (MCQs) covering basics of cell division.
Participants who scored fewer than 4 correct answers were screened out. 
Those who scored 4 or higher proceeded to watch a 15-minute instructional video on the meiosis cell division process and a two-minute video on the AI system they would subsequently utilize for learning.

In the next step, called ``self-learning with AI'', participants were instructed to learn about meiosis cell division using only the assigned system, guided by three predefined learning objectives, and informed that they could use the provided notepad (described below) for note-taking to support their learning. The system interface displayed two tabs for the \DocSearch~condition and three tabs for \TexDoC~and \MuDoC.
The first `Objectives' tab, in all conditions, presented the learning objectives. 
For the \TexDoC~and \MuDoC~conditions, the middle `Chat' tab contained the chat interface.
The final `Document' tab, in all conditions, displayed the source document, the OpenStax Biology textbook \cite{julianne_zedalis_biology_2018} (Chapter 4-14, 447 pages\footref{footnote:supplementary-material}), and a search bar in the case of \DocSearch.
In all conditions, the notepad was displayed on the right side, occupying one-fourth of the interface width.
The interface required participants to spend at least 15 minutes (by disabling the `Next' button) and at most 25 minutes (by automatically advancing the session) on self-learning.
Snapshots from the self-learning session are shown in Figure \ref{fig:learning-session}.

Next, participants completed a post-test with 10 MCQs related to the three learning objectives.
Unlike the pre-test which focused on broader knowledge of cell division with 3 visual-based questions, the post-test focused on three specified learning objectives related to meiosis cell division with 7 visual-based questions.
Finally, they answered survey questions with Likert-scale responses (see Figure \ref{fig:learning-experience}), provided subjective feedback and answered demographic questions.
The survey, learning objectives, pre- and post-tests, and other materials are provided in the supplementary material\footref{footnote:supplementary-material}.

\vspace{3pt}
\textbf{Data Collection:} In addition to test scores and surveys, we measured the time spent using the assigned system, the number of queries, and the frequency of adding or editing notes in the provided notepad across all three conditions (Table \ref{tab:test-scores}). 
For \MuDoC~and \TexDoC, we also measured the average response length from participants' conversations, usage of interactive citations, and the proportion of time spent on the textbook tab (Section \ref{subsec:MuDoCvsTexDoC}).

\section{Results}
\subsection{Pre-test and Post-test Scores}

\begin{table}[t!]
    \centering
    \caption{M, SD values for \textbf{Pre-}, \textbf{Post-test} scores (out of 10), \textbf{Time} (mins) spent interacting with AI, number of \textbf{Queries} and edits to \textbf{Notes} (highest in bold).}
    \setlength{\tabcolsep}{3.5pt}
    \begin{tabular}{r c c c c c c }
        \textbf{System} & \textbf{N} & \textbf{Pre-test} & \textbf{Post-test} & \textbf{Time} & \textbf{Queries} & \textbf{Notes}\\
        \hline 
        \MuDoC~& 42 & 6.92 (1.25) & \textbf{7.24} (1.15) & 17.6 (2.9) & 7.26 (3.32) & 7.1 (7.3) \\
        \TexDoC~& 41 & 6.91 (1.32) & 6.55 (1.11) & \textbf{18.0 (3.7)} & \textbf{8.29 (5.02)} & 7.5 (8.2) \\
        \DocSearch~& 41 & \textbf{6.98} (1.29) & 6.77 (0.93) & 17.6 (3.5) & 5.27 (2.74) & \textbf{11.4 (8.4)}

    \end{tabular}
    \label{tab:test-scores}
\end{table}

Descriptive statistics for performance across the three experimental conditions are presented in Table \ref{tab:test-scores}. 
To test analysis of variance (ANOVA) assumptions, normality was assessed using Shapiro-Wilk tests for each group (all $p > 0.05$) and homogeneity of variance was confirmed using Levene's test, $F(2, 121) = 0.117, p = .89 > .05$ for pre-test scores and $F(2, 121) = 1.711, p = .18 > .05$ for post-test scores.
A one-way ANOVA on pre-test scores indicated no difference across the three groups, $F(2, 121) = 0.27, p = .97, \eta^2 < .001$,
confirming that participants had comparable prior knowledge across the three groups.

A one-way ANOVA on post-test scores revealed a statistically significant effect of the AI system type with a moderate effect size, $F(2, 121) = 4.60, p = .012, \eta^2 = 0.071$. 
Post-hoc comparisons using Tukey’s honestly significant difference (HSD) test showed that \MuDoC~participants achieved significantly higher scores than \TexDoC\ ($p = .010$, $diff = 0.69$). 
While \MuDoC~also outperformed \DocSearch~($diff = 0.47$), this difference did not reach the threshold for adjusted significance in the Tukey HSD test ($p = .11$).

We also examined the influence of prior knowledge (pre-test scores) on learning outcomes (post-test scores) using an Ordinary Least Squares (OLS) regression model while controlling for the experimental condition.
As expected, pre-test scores were significantly positively associated with post-test scores ($\beta = 0.21, p = .005$), providing evidence that the tests are sensitive to differences in prior knowledge and learning gains.

\subsection{Why \MuDoC~is better than \TexDoC?}
\label{subsec:MuDoCvsTexDoC}

To confirm that \MuDoC~leads to better performance than \TexDoC~due to its high-quality multimedia content, we performed in-depth analysis of user interactions and system behavior. 
 While none of the behaviors we captured showed a statistically significant difference, \MuDoC~group consistently had lower mean values.
They jumped to the textbook using interactive citations less frequently ($p=.08, d = -0.38$ using t-test), 
asked fewer questions ($p=.27, d=-0.36$), 
spent less time with the AI system ($p = .31, d = -0.11$), 
spent a lower proportion of time on the textbook tab ($p = .60, d = -0.12$), 
and made fewer updates to their notes ($p = .80, d = -0.06$).
In terms of system behavior, the average response length of \MuDoC~(excluding figures) was lower than that of \TexDoC~, but not significantly ($p = 0.10, d = -0.36$).
Besides text, an average conversation with \MuDoC~had $M = 3.1, SD = 1.2$ unique images and a total of $M = 6.2, SD = 2.3$ images.

These results show that there was no significant differences in learning behaviors and system behavior, except for the visuals included in \MuDoC~responses, which likely drove the improvements in learning outcomes.
Further, if we examine the weak trends discussed above, their consistency suggests that \MuDoC~group may have allocated more time to examining the visuals rather than relying on the textbook or asking follow-up questions. 
\TexDoC~users likely had to read complex text, ask follow-up questions, and seek visuals in the source textbook, all of which are time-consuming and cognitively demanding. 

\textit{How does \DocSearch~compare?} Compared to the systems with conversational AI, the \DocSearch~group made significantly fewer queries ($p < .001, d=-0.63$), but made significantly more updates to their notes ($p = .008, d = 0.50$). 
This group made $M=5.27$, $SD=2.74$ queries and examined $M=18.8$, $SD=10.1$ search results.
In summary, \DocSearch~interactions show a more traditional approach to reading with fewer queries and more note-taking.

\subsection{Quantitative Data on Learning Experience}

\begin{figure}[t!]
    \centering
    \includegraphics[width=\linewidth]{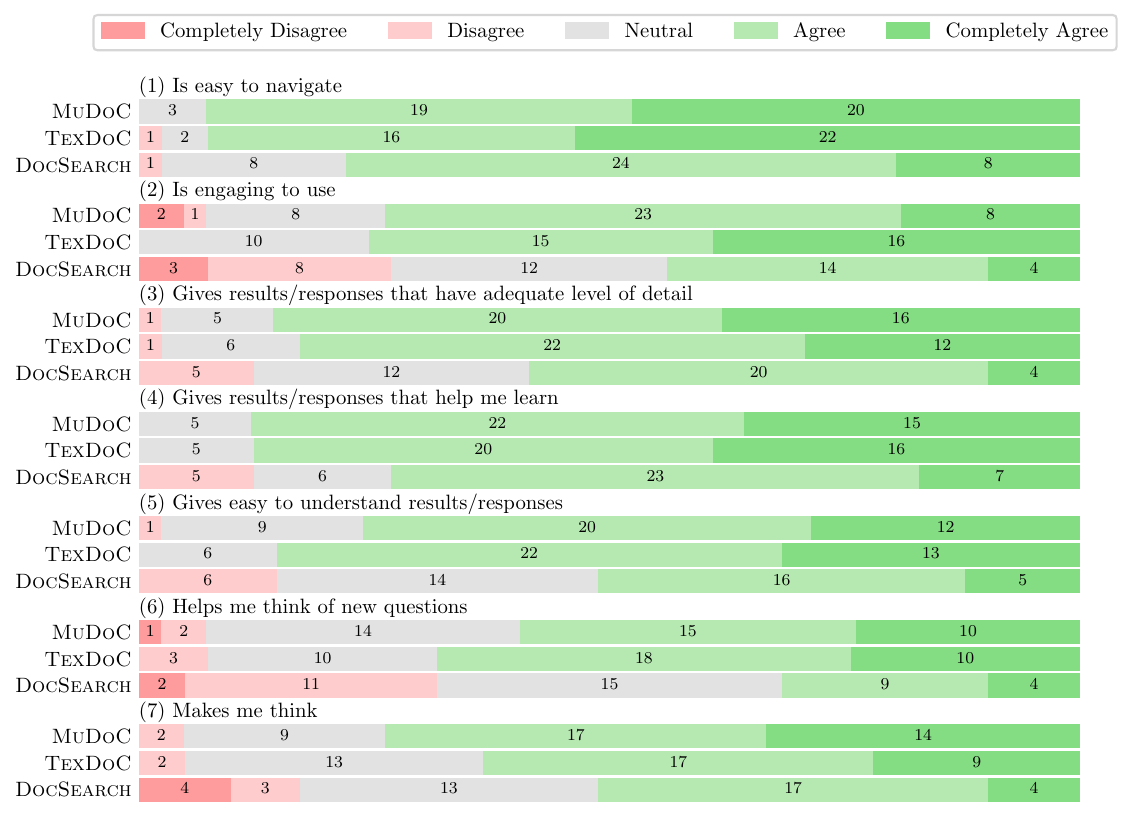}
    \caption{Perceived Learning Experience with \MuDoC, \TexDoC~and \DocSearch.}
    \label{fig:learning-experience}
\end{figure}

The results from seven Likert-scale survey questions are summarized in Figure \ref{fig:learning-experience}. 
Since normality and homogeneity of variance are not satisfied by Likert-scale data, we used the non-parametric Kruskal-Wallis test. 
The results revealed significant between-group differences across all seven questions ($p < .05$), with effect sizes ($\eta^2$) ranging from moderate (.051) to large (.123). 
Post-hoc analyses were performed using Dunn's test with Holm correction for multiple comparisons. 

\textit{Usability} (1, 2 in Figure \ref{fig:learning-experience}):
The analysis indicates that conversational interfaces (\MuDoC~and \TexDoC) provided a superior user experience compared to \DocSearch. 
Participants found the two systems significantly easier to navigate ($p = .002,  \eta^2 = .09$) and more engaging to use ($p < .001, \eta^2 = .12$). 
No significant differences between \MuDoC~and \TexDoC~suggest that conversationality was the primary driver of better usability ratings.

\textit{Response Quality} (3, 4, 5 in Figure \ref{fig:learning-experience}):
In terms of perceived quality, both conversational AI systems outperformed \DocSearch~in providing detailed ($p = .001, \eta^2 = .10$), helpful ($p = .017, \eta^2 = .05$), and clear responses ($p = .001, \eta^2 = 0.10$).
Again, we did not see a difference between \MuDoC~and \TexDoC, suggesting that conversationality in AI systems leads to better perceived quality for information needs as compared to stateless semantic search in \DocSearch.

\textit{Cognitive Engagement} (6, 7 in Figure \ref{fig:learning-experience}):
Both conversational systems significantly outperformed the baseline in helping users formulate follow-up questions ($p = .001, \eta^2 = .10$), indicating a shift from passive reading to active cognitive engagement. 
However, for the question `makes me think' ($p = .012, \eta^2 = .06$), \MuDoC~was the only system to maintain a statistically significant advantage over the \DocSearch~baseline ($p = .010$ in Dunn's test). 
\TexDoC~showed a positive trend but did not reach statistical significance ($p = .163$ in Dunn's test). 
This suggests that the integration of visuals, unique to \MuDoC, helps learners integrate information from two modalities, therefore making them think more.

\subsection{Qualitative Data on Learning Experience}

Participants answered two survey questions about the strengths and limitations of the AI systems they were assigned.
We split their responses into distinct assertions and thematically organized similar assertions, of which the most common ones are elaborated below. 

\paragraph{What was the best part of using `System X' during the self-learning session?}
\DocSearch~was primarily lauded by 18 participants for its \textit{efficiency}, specifically its ability to act as a `smart' search tool; 
one participant noted that it ``likely saved 20-30 minutes of reading'' by streamlining the identification of relevant material in the textbook. 
In contrast, \TexDoC's main strength was in its \textit{transparency} (14 participants), with users appreciating the ``links back to where in the textbook [it] was referencing,'' which served as a vital tool for information verification, corroborating findings from MuDoC 1.0 \cite{taneja_towards_2025}. 
While \MuDoC~had identical source attribution, these strengths were overshadowed by participants' appreciation for its \textit{multimodal grounding} and \textit{simplified responses} (9+9 participants). 
Users highlighted that the system ``integrated visual aids to help explain the concepts'' while ``summariz[ing] text that is typically long and tiring.''

\paragraph{What are some frustrations you experienced when using `System X'?}
\DocSearch~users reported significant frustration with its perceived lack of `intelligence', with 10 participants noting it felt like a `glorified search tool' rather than a helpful AI. 
In the \TexDoC~condition, the most prominent frustration was the \textit{lack of visual aids} (9 participants), which forced users to ``switch back and forth between tabs'' to view necessary diagrams. 
While \MuDoC~successfully addressed the visual needs, a few participants mentioned limitations such as \textit{cognitive overload} (7 participants) and \textit{system latency} (6 participants) that were also brought up by 5 and 3 \TexDoC~participants respectively, while only 4 \DocSearch~participants indicated the former as their biggest frustration.
Very few participants in each group (1-2) reported difficulties with the document or chat interface, which alleviates concern about the interface design negatively impacting the learning experience.

\paragraph{Summary:} 
For \DocSearch, across the two responses, 13 participants expressed a desire for content summarization, reflecting the prevalent expectation of synthesized summaries from AI systems.
For \TexDoC, the feedback centered on the limitations of a text-only conversational AI, particularly for the complex STEM topic they were learning.
\MuDoC~received the most enthusiastic feedback overall, particularly for learning ``without needing to look at paragraphs of text'' and ``the fact [that] it pulled [visuals] from the textbook''.

\section{Discussion and Future Work}
\label{sec:discussion}

The results of our study suggest a complex interplay between multimodality, conversationality, learning experience, and learning outcomes.

\MuDoC~enabled conversational multimedia learning that led to significantly better \textit{learning outcomes} with a moderate effect size compared to \TexDoC. It also led to better learning outcomes compared to \DocSearch~, although this result was not statistically significant.
In terms of \textit{learning experience}, including ease of use, engagement, and perceived helpfulness, \MuDoC~is at par with \TexDoC~and outperforms \DocSearch.
Loosely speaking, trends show that \MuDoC~> \DocSearch~$\geq$ \TexDoC~for learning outcomes (Table \ref{tab:test-scores}), while \MuDoC~$=$ \TexDoC~> \DocSearch~for learning experience (Figure \ref{fig:learning-experience}).
We interpret these results through the lens of Cognitive Load Theory (CLT) \cite{sweller_cognitive_2011}.

According to CLT, effective learning occurs when instructional design reduces extraneous cognitive load and promotes germane load within the limits of working memory \cite{sweller_cognitive_2011}.
\textit{Extraneous load} is the mental effort caused by poorly designed learning materials or distracting environmental factors, while 
\textit{germane load} is the beneficial mental effort used to process new information, build mental models, and integrate knowledge into long-term memory.

Multimedia learning leads to higher germane load through careful content design \cite{mayer_multimedia_2002}.
In the case of \DocSearch, while the presence of multimedia content would have led to high germane load, lack of conversationality likely increased extraneous load---by requiring learners to repeatedly search and integrate fragmented content---potentially exceeding cognitive capacity. 
This imbalance may explain the poorer learning experience observed with \DocSearch.
In the case of \TexDoC, low extraneous load due to conversationality likely led to better learning experience, but lower germane load due to lack of multimodality could have resulted in poorer learning outcomes.
Compared to \TexDoC, in \DocSearch, in addition to the multimodal content, the more frequent note-taking (Table \ref{tab:test-scores}) would have also contributed to higher germane load. This may explain why, despite higher extraneous load and a poorer learning experience, the \DocSearch\ group's learning outcomes were comparable to \TexDoC.
Finally, in the case of \MuDoC, conversationality reduced extraneous load, thereby enhancing the learning experience. 
This preserved cognitive capacity could have enabled learners to capitalize on the heightened germane load due to multimodality, leading to better learning outcomes.
In summary, these findings suggest that 
\textit{
conversationality reduces extraneous load by maintaining contextual continuity, eliminating the need for repeated search and simplifying the integration of information. 
Further, conversational multimedia learning exhibits an improved cognitive load balance by lowering extraneous load and increasing germane load, leading to both improved learning outcomes and experience.
}

It is also worth noting that there is a striking disconnect between participants' perceptions of \TexDoC\ and its actual impact on learning outcomes---participants rated \TexDoC~as highly engaging and easy to use, yet it yielded the lowest post-test scores. 
While many in the \TexDoC~group realized that visuals would support learning, they did not spend significantly more time exploring visuals in the textbook compared to the \MuDoC~group, likely as that would have led to increased extraneous load \cite{kool_decision_2010}.
In contrast, by reducing the effort required to access and interpret visuals, \MuDoC\ facilitated integration between text and graphical visuals, therefore replacing extraneous load with germane load.
This suggests that \textit{when reduced extraneous load from conversationality is not complemented by higher germane load, it can lead to a \textit{fluency effect} where ease of digesting information is mistaken as a sign of learning} \cite{oppenheimer_secret_2008}.

Finally, \MuDoC~establishes the efficacy of recent MLLMs to adaptively generate high-quality multimedia content for educational contexts. 
Further improvements in image retrieval, image understanding in MLLMs, and generative AI for diagram creation and editing will also enhance context-specificity of visuals. 
Our results also highlight opportunities for future research on how conversational AI can be effectively designed to enhance learning. 
While conversational interaction appears beneficial for improving subjective learning experience, text-only conversational systems may be insufficient for improving learning outcomes. 
In visually rich STEM domains, the inclusion of relevant textbook-based visuals alongside textual explanations is both necessary and beneficial for learning, as evidenced by our findings. 
In domains where visuals are less critical, future work should explore alternative design strategies for introducing desirable germane load---such as metacognitive prompting \cite{singh2025enhancing} or introducing “friction” in human-AI interactions \cite{kazemitabaar2025exploring}---within conversational AI systems to support learning outcomes.

\textit{Limitations}:
First, the study was conducted within a single domain and for a single biology topic. 
Second, the time constraints (15-25 minutes) may have affected learning outcomes, but this effect should be minimal as 88\% of participants completed the learning session before the time limit, suggesting that the time allotted for the task was sufficient.
Third, we assessed learning immediately after the intervention; future work should also consider measuring impact on long-term learning outcomes such as retention.
Additionally, we grounded our interpretation in CLT, but did not directly measure cognitive load during learning. Future work should include explicit cognitive load measures \cite{korbach_measurement_2017} to better understand the influence of conversational multimedia learning.
Fourth, the study recruited participants via Prolific. 
Future research should examine these findings in more authentic educational settings.
Finally, \MuDoC's performance may be dependent on the quality of the source textbook and visuals which impacts the accuracy of the document preprocessing pipeline, a subject of system evaluation that is outside the scope of this work.

\section{Conclusion}
\label{sec:conclusion}

In this paper, we described \MuDoC, a conversational multimedia learning system that generates grounded, interleaved text-and-image responses based on textbook content, 
and compared it to two baselines---\TexDoC~({\Tex}t-only Mu-\DoC) and \DocSearch~({\Doc}ument with semantic \Search)---through an RCT with $N=124$ participants. 
\MuDoC~led to the highest post-test scores and a better learning experience based on qualitative and quantitative feedback, showing that multimodal LLMs can be effectively leveraged for learning by introducing interleaved responses containing grounded visuals.
Examining the results through the lens of Cognitive Load Theory (CLT), we argued that \textit{conversational multimedia learning} improves cognitive load balance by reducing extraneous load as a result of conversationality while increasing germane load as a result of multimodality. 
Our findings also suggest that conversational AI systems can lead to an illusion of learning, unless complemented by strategies to cultivate higher germane load, such as visually-enriched explanations. 
Future work should explore the application of conversational multimedia learning across diverse domains and develop improved methods for image retrieval, understanding, and context-specific visual generation to further enhance instruction.

\begin{credits}
\subsubsection{\ackname} 
We are grateful for the support provided by National Science Foundation under Grant No. 2247790 and Grant No. 2112532. 
We also wish to thank Dr. Emily G. Weigel and Joon Kum for support in creation of assessments. 
\end{credits}

\bibliographystyle{splncs04}
\bibliography{references,references2}

\end{document}